# Singular states of resonant nanophotonic lattices

Y. H. Ko, K. J. Lee, F. A. Simlan and R. Magnusson*

*Department of Electrical Engineering, University of Texas at Arlington, Arlington, Texas 76019, USA*
*magnusson@uta.edu

**Fundamental effects in nanophotonic resonance systems focused on singular states and their properties are presented. Strongly related to lattice geometry and material composition, there appear resonant bright chan-nels and non-resonant dark channels in the spectra. The bright state corresponds to high reflectivity guided-mode resonance (GMR) whereas the dark channel represents a bound state in the continuum (BIC). Even in simple systems, singular states with tunable bandwidth appear as isolated spectral lines that are widely sep-arated from other resonance features. Under moderate lattice modulation, there ensues leaky-band meta-morphosis, merging modal bands and resulting in offset dark states and reflective BICs along with transmis-sive BICs within a high-reflectance wideband. Rytov-type effective medium theory (EMT) is shown to be a powerful means to describe, formulate, and understand the collective GMR/BIC fundamentals in resonant photonic systems. Particularly, the discarded Rytov analytical solution for asymmetric fields is shown here to predict the dark BIC states essentially exactly for considerable modulation levels. The propagation con-stant of an equivalent EMT homogeneous film provides a quantitative evaluation of the eminent, oft-cited embedded BIC eigenvalue. The work concludes with experimental verification of key effects.**

## 1. INTRODUCTION

Understanding of wave propagation in periodic systems is foundational for their utility in science and engineering. Thus, the properties of solid-state materials are explained with band theory modeling propagation of electron waves in periodic crystal lattices. Elementary Bragg diffraction reveals energy bands and band gaps and classification of materials as insulators and conductors [1]. Similar bands appear in three-dimensional (3D) dielectric lattices called photonic crystals [2]. The band structure determines how photon propagation is affected by frequency, polarization, and direction. It may be represented in the first Brillouin zone with the first and higher band gaps corresponding to Bragg reflections at increasing frequency [3]. Whereas 3D dielectric periodicity is challenging in experimental realization, there is much current interest in practical film-based 1D and 2D optical lattices with straightforward fabrication. Key physical properties of these elements are explained in terms of the structure of the second (leaky) photonic stopband and its relation to the symmetry of the periodic profile.

When the lattice is confined to a layer thereby forming a periodic waveguide, an incident optical wave may undergo a guided-mode resonance (GMR) on coupling to a leaky eigenmode of the layer system [4-7]. Figure 1(a) models the simplest resonance system possible, namely a subwavelength 1D periodic lattice or grating. Under normal incidence, counter-propagating leaky modes form a standing wave in the lattice. As the modes interact with the lattice, they reradiate [8]. A schematic dispersion diagram is shown in Fig. 1(b). The device works in the second stop band corresponding to the second-order lattice [9]. A given evanescent diffraction order can excite not just one but several leaky modes. To emphasize this point, in Fig. 1(b) we show the stop bands for the first two TE modes. At each stop band, a resonance is generated as denoted in Fig. 1(b). The fields radiated by these leaky modes in a lattice with a symmetric profile can be in phase or out of phase at the edges of the band [10]. At one edge, there is a zero-phase difference, and hence the radiation is enhanced (GMR) while at the other edge, there is a $\pi$ phase difference inhibiting the radiation. In this case, if $\beta = \beta_R + i\beta_I$ is the complex propagation constant of the leaky mode, $\beta_I = 0$ at one edge, which implies that no leakage is possible at that edge marking the condition as a bound state in the continuum (BIC). For asymmetric lattice profiles, guided-mode resonance prevails at each band edge. Fundamentally, one-dimensional

(1D) and 2D resonant lattices operate similarly and thus the description in Fig. 1 applies generally.

Within the domain of nanophotonics, especially relating to new developments in metamaterials and metasurfaces, the resonance phenomena explained with Fig. 1 are of major interest. The leaky (GMR) edge and the nonleaky (BIC) edge are inherent in resonance systems in this class. Historically, BICs were proposed in hypothetical quantum systems by von Neumann and Wigner [11]. In such systems, a completely bound state exists at an energy level above the lowest continuum level. Whereas the term BIC appeared in photonics in 2008 [12], the underlying concept was apparently first reported by Kazarinov et al. in 1976 [13]. These researchers derived a formula for the quality factor of a corrugated waveguide and reported zero radiation loss at the upper band edge when the second-order Bragg condition was satisfied. In a later paper, these effects were elaborated with improved clarity [10]. Analyzing the second-order stop bands, Vincent and Neviere numerically demonstrated the existence of a non-leaky edge pertinent to symmetric gratings whereas asymmetric grating profiles yielded leaky radiant modes at both band edges [4]. Ding and Magnusson manipulated the separation of the non-degenerate leaky resonances associated with asymmetric profiles to engineer the resonant spectral response of periodic films [14]. Experimentally, the nonleaky edge was brought into view in 1998 by imposing asymmetry on an otherwise symmetric 1D periodic structure by variation of the incidence angle [15]; at the time the BIC terminology was not in use.

Reviewing briefly recent works, Marinica et al. proposed a symmetric double-grating structure to support embedded photonic bound states by coupling between two identical resonant grating layers [12]. Hsu et al. experimentally showed a diverging radiation Q factor as a signature of embedded bound states in a 2D modulated layer of silicon nitride [16]. By tuning the structural symmetry or coupling strength between different resonance channels, quasi-BICs can be generated possessing ultranarrow linewidths [17-19]. Such high-Q resonances neighbouring BIC points enable ultra-sharp transmission and reflection spectra, yielding giant near-field enhancement and various promising applications including BIC-based chirality [20-22], lasing [23-25], nonlinearity [26-29], modulation [30] and sensing in various spectral regions [31]. In addition to the symmetry protected BICs in the Brillouin zone center at a $\Gamma$ point, there exist off-$\Gamma$ BIC states under non-normal incidence, sometimes called accidental BICs or quasi-BICs [32-38]; such BIC states provide additional degrees of freedom and application possibilities.

In this paper, we treat leaky band metamorphosois where evolution of the leaky band structure with lattice modulation strength is evaluated. The resulting singular state is then shown to emerge as an isolated resonance feature brought to perfect narrowband reflection under broken symmetry. Effective-medium theory (EMT) based on the Rytov symmetric and asymmetric formalisms is shown to model the resonant and BIC states with high precision. The veracity of the Rytov EMT in delivering homogeneous waveguide slabs that contain the GMR/BIC resonance spectral properties is shown for practical modulation levels up to 3. Experimental results support the main theoretical conclusions.

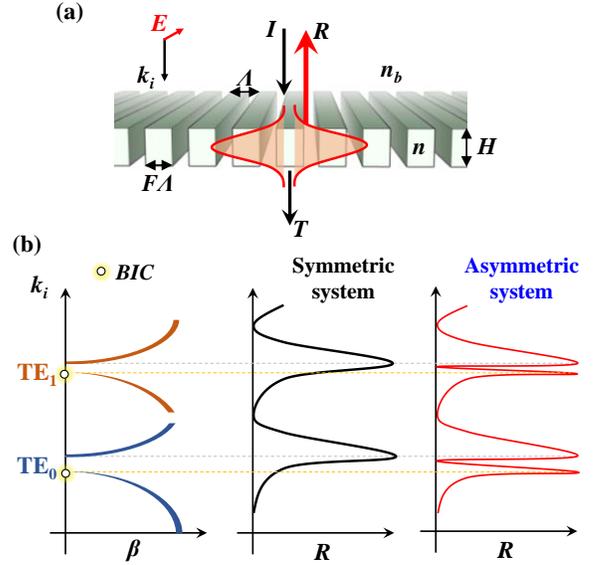

**Fig. 1.** (a) A schematic view of the simplest subwavelength resonance system. The model lattice has thickness ($H$), fill factor ($F$), period ($\Lambda$), and refractive indices of background and lattice material ($n_b, n$). When phase matching occurs between evanescent diffraction orders and a waveguide mode, a guided-mode resonance occurs. $I$, $R$, and $T$ denote the incident wave with wavelength $\lambda$, reflectance, and transmittance, respectively. (b) A schematic dispersion diagram of a resonant lattice at the second stop band. For a symmetric lattice, the leaky edge supports guided-mode resonant radiation while the non-leaky edge hosts a non-radiant bound state. This picture applies to both TE (electric field vector normal to the plane of incidence and pointing along the grating grooves) and TM (magnetic field vector normal to the plane of incidence) polarization states. Here, the grating vector has magnitude $K = 2\pi/\Lambda$, $k_i = 2\pi/\lambda$, and $\beta$ denotes a propagation constant of a leaky mode.

## 2. LEAKY-BAND METAMORPHOSIS

As in Fig. 1(b) pertaining to a symmetric lattice, each resonant mode has a clear GMR leaky edge and a corresponding BIC non-leaky edge. This always holds for "weakly" modulated lattices. On increase of modulation, the band deviates and the GMR-BIC pairing is obscured. Key aspects of the band transformation can be brought out via the simple model in Fig. 2(a). We compute incidence-angle ($\theta$) dependent zero-order reflectance ($R_0$) in TE polarization. Figure 2(b) displays the $R_0(\theta, \lambda)$ map for $n$=1.4 where the Rayleigh lines ($\lambda_R = \Lambda \pm \sin\theta$) are marked by white dashed lines. At normal incidence, a single $R_0$ peak appears at the upper band edge generated by symmetry-allowed GMR marked as s-GMR with the attendant leaky mode having a $TE_0$ mode shape. At nonnormal incidence under asymmetry, new $R_0$ peak branches form in the lower band denoted a-GMR also having a $TE_0$ mode character consistent with Fig. 1(b) in low modulation. On approaching normal incidence and symmetry, the a-GMR vanishes becoming a BIC. Upon increasing n to 1.7, in Fig. 2(c), the bandwidth of the high $R_0$ peak expands due to the strengthened dielectric modulation ($\Delta\varepsilon=n^2-n_a^2$). Particularly, in the upper band, the high $R_0$ region is enlarged around the s-GMR within the $\lambda_R$ lines achieving wideband reflection. The leaky modes forming the contiguous high-reflection region at the upper band now have mixed $TE_0$ and $TE_1$ shapes while the lower band retains its $TE_0$ mode character. Beyond the $\lambda_R$ lines, $R_0$ rapidly decreases because of leakage to higher-order reflection and transmission. Concomitantly, the band gap between the s-GMR and BIC states increases. In a prior study [39], we explain how the increase of $\Delta\varepsilon$ opens the band gap after band closure affected by the first and second Fourier harmonics of the dielectric function in the small-modulation limit. In the current example, the band gap closes near $n$=1.1. On setting $n$=2, in Fig. 2(d), the reflectance bandwidth broadens further, and the band gap increases. Remarkably, the BIC state separates widely from other prominent resonance features and resides in an extensive, low-reflectance region. At θ=0, the fundamental $TE_0$ mode is missing. Setting the angle to a small value brings out a sharp reflective resonance with $R_0$=1 mediated by the fundamental, now leaky, mode. The narrow-line resonance emerging from a dark background appears as a singular state in the spectrum. As an aside, in the upper band of Fig 2(d), a small incident angle induces a sharp resonant dip in the background of high reflection. Recently, such zero-order transmittance ($T_0$) spectra have been connected to electromagnetically induced transparency (EIT) as being EIT-like or EIT-analogous.

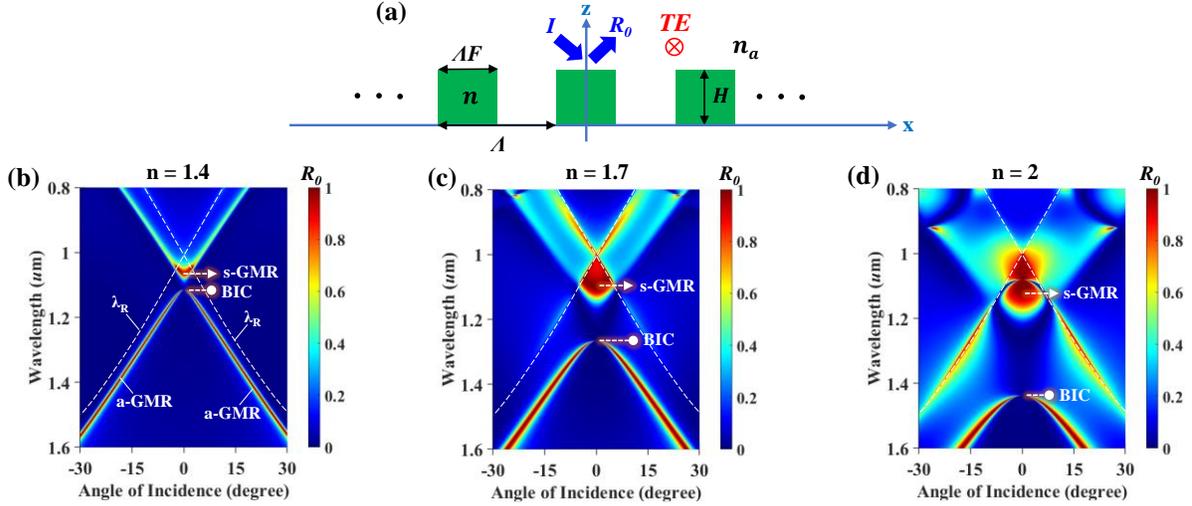

**Fig. 2.** (a) Modeling a simple subwavelength photonic lattice in air ($n_a$=1). Parameters are period ($\Lambda$=1 µm), grating width ($W=\Lambda F$=0.5 µm) and height ($H$=0.5 µm) under TE-polarized plane-wave incidence. Angular zeroth-order reflectance $R_0(\theta,\lambda)$ properties are analyzed with rigorous coupled wave analysis (RCWA) for lattice refractive indices (b) $n$ =1.4, (c) $n$=1.7 and (d) $n$=2. Note that we use an inverted wavelength scale for consistency with proper band terminology where the upper band refers to high frequency. The Rayleigh wavelength ($\lambda_R$) is denoted by white dashed lines.

## 3. THE SINGULAR STATE AND ITS EIGENVALUE

We now analyze the singular state in detail. We show that it is possible to establish a homogeneous equivalent waveguide representing the resonant lattice nearly exactly. The propagation constant $\beta$ of the leaky mode is then known and represents the embedded eigenvalue belonging to the unexcited bound state.

Figure 3(a) shows mode loci for the slab waveguide modes supported in the equivalent 1D lattice with refractive index n=2 as displayed in Fig. 2(a). To define the equivalent slab, the periodic layer is homogenized by effective medium theory (EMT) with the Rytov formalism in TE polarization [40, 41] expressed as

$$\sqrt{1-(n_{EMT}^{TE})^2}\tan[\tfrac{\pi\Lambda}{\lambda}(n_a^2-F)\sqrt{n_a^2-(n_{EMT}^{TE})^2}] = \\ -\sqrt{n^2-(n_{EMT}^{TE})^2}\tan[\tfrac{\pi\Lambda}{\lambda}F\sqrt{n^2-(n_{EMT}^{TE})^2}] \quad \textbf{(1)}$$

We use the first-order solution $n_1^{TE}(\lambda,\Lambda)$ as an effective refractive index found by solving Eq. (1) corresponding to excitation by first-order diffracted waves. In his original 1956 paper, Rytov provided two solutions where one corresponded to complete symmetry represented by Eq. (1) of the local fields and the other to asymmetric fields [40]. He discarded the asymmetric solution as being unphysical and of no interest. But most importantly, the asymmetric local fields actually correspond to BIC states with the Rytov solution describing them with high precision as we now show. The discarded solution is [40]

$$\sqrt{1-(v_{EMT}^{TE})^2}\tan[\tfrac{\pi\Lambda}{\lambda}F\sqrt{n^2-(v_{EMT}^{TE})^2}] = \\ -\sqrt{n^2-(v_{EMT}^{TE})^2}\tan[\tfrac{\pi\Lambda}{\lambda}(n_a^2-F)\sqrt{n_a^2-(n_{EMT}^{TE})^2}] \quad \textbf{(2)}$$

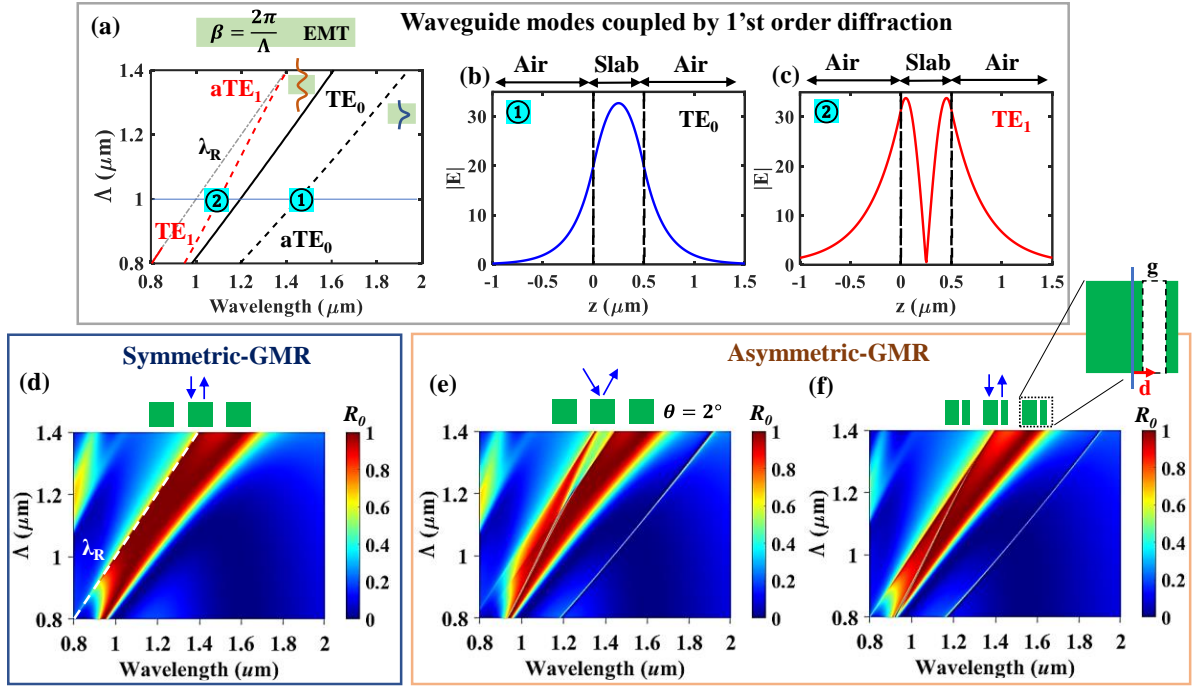

**Fig. 3.** Analysis of phase matching to resonant states in symmetric and asymmetric optical lattices with n=2. (a) Mode loci in the equivalent lattice homogenized by the Rytov EMT and thus becoming a slab waveguide. At points of ① and ②, with Λ=1 μm, transverse profile of |E| is calculated showing (b) $TE_0$ and (c) $TE_1$ mode shapes. The $R_0(\lambda, \Lambda)$ maps for (d) $\theta$=0°, (e) $\theta$=2° and (f) asymmetric geometry at $\theta$=0° exhibit s-GMR and a-GMR features. In (f), the square rod is sliced by an air gap (g=0.05 μm) where the distance between centers of rod and air gap is d=0.05 μm.

From Eq. (2), we find $v_1^{TE}(\lambda, \Lambda)$ corresponding to the effective index in the asymmetric case under leaky-mode excitation by the first evanescent diffraction order and use it to establish the equivalent homogeneous film. Then, we solve the classic slab waveguide eigenvalue problem for the q-th TE mode using $n_1^{TE}$ and $v_1^{TE}$ [42]

$$tan\left(\frac{n_1^{TE} k_i H}{2} - \frac{\pi q}{2}\right) = \frac{\sqrt{(\beta)^2 - (n_a k_i)^2}}{n_1^{TE}} \quad \text{(Symmetric)} \quad (3)$$

$$tan\left(\frac{v_1^{TE} k_i H}{2} - \frac{\pi q}{2}\right) = \frac{\sqrt{(\beta)^2 - (n_a k_i)^2}}{v_1^{TE}} \quad \text{(Asymmetric)} \quad (4)$$

Here, in the slab, the propagation constant ($\beta$) equals the grating vector ($K = 2\pi / \Lambda$) under input wavevector ($k_i = 2\pi / \lambda$) on phase-matching to the first evanescent diffraction order.

In Fig. 3(a) the solid lines show the mode loci for the symmetric situation found by Eqs. (1) and (3). In the plot, there are two TE modes under or near the Rayleigh $\Lambda=\lambda_R$ line where $TE_0$ and $TE_1$ denote the fundamental (q=0) and first-order (q=1) waveguide modes. For each mode, as shown in illustrative insets, the local field intensity is confined in a single (at film center) and double peak (near film edges) as typical in symmetric slab waveguides. The asymmetric GMR modes similarly identified by solving Eqs. (2) and (4) in the slab waveguide are displayed by dashed lines in Fig. 3(a). As explained in detail in [41] for guided-mode resonant lattices, operating with first-order diffraction (m=1), the lattice is homogenized by vertical indices $n_1^{TE}$ and $v_1^{TE}$ with corresponding wave numbers $k_i n_1^{TE}$ and $k_i v_1^{TE}$ whereas lateral modes see $\beta=K$.

Figures 3(b) and 3(c) show simulated magnitude distribution of the electric field (|E|) of $TE_0$ and $TE_1$ modes at $\Lambda$=1 μm where the corresponding $\lambda$ locates at ① 1.531 μm and ② 1.153 μm. Figure 3(d) shows the $R_0(\lambda, \Lambda)$ map at normal incidence as computed exactly with rigorous coupled-wave analysis (RCWA) [43, 44]. Bounded by the $\lambda_R$ line, a high $R_0$ band forms around the $TE_1$ and $TE_0$ mode lines in Fig. 3(a). Thus, the characteristics and mode support picture of the EMT slab guide agree well with the rigorous numerical results. Figure 3(e) shows the $R_0(\lambda, \Lambda)$ map for the asymmetric case with $\theta$=2°. The singular state locus is seen as a narrow high-reflection band offset from the other resonance features by $\lambda \sim$200 nm. It is predicted by the EMT model nearly exactly as seen by comparing with the $aTE_0$ locus in Fig. 3(a). Similarly, the resonance line marked $aTE_1$ agrees well with the numerical map.

An alternate approach is to break the symmetry geometrically while maintaining normal incidence as noted in Fig. 3(f) where similar spectra appear. To implement the asymmetric geometry, each rod is asymmetrically sliced by an air gap (g=0.05 μm) where the cutting position (d=0.05 μm) is decided by the distance between the center of the ridge and the air gap. In the $R_0(\lambda, \Lambda)$ maps, sharp resonances appear near the modal curves of Fig. 3(a) and generally exhibit a Fano line shape. As the a-GMR is induced in broken symmetry under phase-matching, there open radiation channels with pathways in both transmission and reflection. The channel at the aTE$_0$ mode location produces a narrow-band perfect reflection $R_0=1$ whereas the one at aTE$_1$ appears in the wide-reflectance band and produces a sharp transmission $T_0=1$ peak or "BIC-EIT." Comparing Fig. 3(a) to Figs. 3(e) and 3(f) shows, quite surprisingly, that the discarded Rytov solution in Eq. (2) predicts the BIC state essentially exactly. The embedded eigenvalue of the aTE$_0$ mode corresponding to the reflective BIC state is $\beta(\lambda)=K=2\pi/\Lambda$ and is thus given numerically in Fig. 3(a).

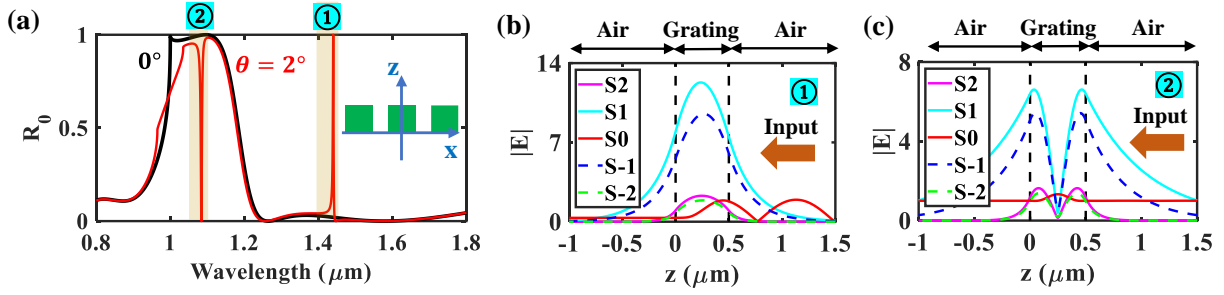

**Fig. 4.** Resonance spectra and leaky-mode profiles under a-GMR. (a) $R_0$ spectra at $\theta=0°$ and $\theta=2°$ of the model 1D optical lattice {$\Lambda=1$ μm, $F=0.5$, $H=0.5$ μm, $n=2$} at a crosscut of Figs. 4(d) and 4(e) at $\Lambda=1$ μm. For a-GMR reflectance peak and null at ① and ②, the corresponding leaky-mode profiles are presented in (b) and (c).

Figure 4 verifies the a-GMR spectral identity. In Fig. 4(a), two $R_0$ spectra are given by horizontal crosscuts of Figs. 3(d) and 3(e) at $\Lambda=1$ μm. In the plot, at $\theta=2°$, two sharp peaks locate at ① $\lambda=1.441$ μm and ② $1.084$ μm in the broken symmetry. For each position, corresponding leaky-mode field profiles are characterized by RCWA in Figs. 4(b) and 4(c) where $S_{\pm m}$ denote amplitudes of the coupled $\pm m$-th diffraction orders. Comparing to Figs. 3(b) and 3(c), we see that the modes driven by $S_{\pm 1}$ and $S_{\pm 2}$ match the reference TE$_0$ and TE$_1$ mode profiles, proving that the diffraction orders are coupled to the leaky waveguide modes in broken symmetry. Furthermore, these modes generate the attendant radiation channels. As seen in Fig. 4(b), the $S_0$ is reflected along the +Z direction (i.e., $R_0$) by the resonant interaction of the incident light with the TE$_0$ mode of the $S_{\pm 1}$ and $S_{\pm 2}$. On the other hand, in Fig. 4(c), the $S_0$ passes through the lattice (i.e., $T_0$) after interfering with the TE$_1$ mode.

## 4. VERACITY OF RYTOV'S SOLUTIONS IN $\Delta\varepsilon$

Figure 5 shows the band progression versus the refractive index modulation strength $\Delta\varepsilon = n_H^2 - n_L^2$ where the $n_H$ and $n_L$ are high and low refractive indices of the lattice. In Fig. 5(a), the numerical $R_0$ loci at $\theta=0°$ appear as s-GMR bands with TE$_0$ and TE$_1$ being the radiative leaky modes. The dark states can be tracked by slightly breaking the symmetry as shown in Fig. 5(b). At $\theta$ =0.1°, two dark states are seen to gradually move to lower energy when $\Delta\varepsilon$ increases. One dark state remains in a bound state at longer wavelengths while the other locates in the high reflection band of the s-GMR $TE_0$ and $TE_1$ modes. Upon breaking the symmetry, in relatively strong modulation, each dark state is transferred to a narrow reflective GMR and EIT-GMR, respectively. Figure 5(c) confirms the veracity of the formulation of s- and a-GMRs with symmetric and asymmetric Rytov-type EMT. Each modal line in Fig. 5(c) matches the numerical resonant leaky bands nearly perfectly.

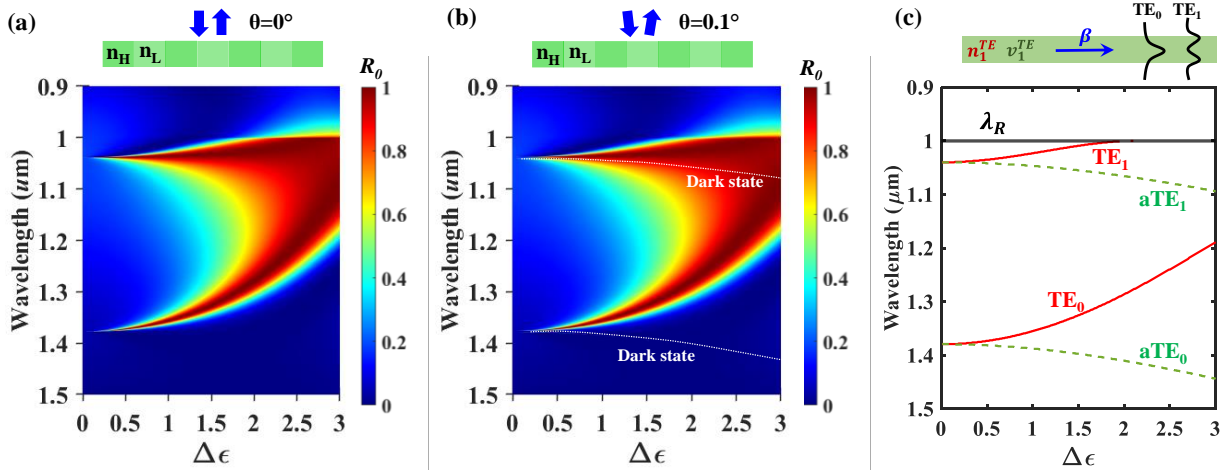

**Fig. 5.** Band progression relative to index modulation strength $\Delta\varepsilon = n_H^2 - n_L^2$. The lattice has parameters ($\Lambda=1$ μm, $F=0.5$ and $H=0.5$ μm) while keeping a fixed average refractive index ($n_{avg} = \sqrt{Fn_H^2 - (1-F)n_L^2}$). Reflectance $R_0(\Delta\varepsilon, \lambda)$ is found by RCWA at (a) normal incidence $\theta=0°$ and (b) off-normal incidence $\theta=0.1°$. In the symmetric system of (a), two (upper) modal bands appear. In (b), at slightly broken symmetry, two hidden dark states appear as lower bands. (c) Mode loci of the equivalent TE slab applying the symmetric and asymmetric Rytov EMT where the $\lambda_R$ indicates the Rayleigh wavelength. As $\Delta\varepsilon$ increases, the pair of bands split in symmetric and asymmetric GMRs, with an excellent match to (b). Interestingly, the lower band of the aTE$_0$ mode is highly separated from the upper band of the TE$_0$ mode, which leads to the isolated singular states of central importance.

## 5. EXPERIMENTAL RESULTS

The singular state is demonstrated by a silicon (Si) 1D lattice structure as shown in Fig. 6(a). For facile device fabrication, avoiding the membranes in air presented in the theory sections above, we use crystalline Si (c-Si, $n_{Si}$=3.48) on a quartz ($n_{Qz}$=1.45) substrate and index matching oil ($n_{oil}$=1.45) to reach vertical symmetry. After cleaning the SOQ wafer with 620-nm-thick c-Si layer (Shin-Etsu Chemical, Co., Ltd.), we adjust the thickness of the c-Si layer by dry etching. Then, 1D c-Si grating patterns are prepared by UV laser interference lithography and a dry etching process [45]. Illuminating a coherent beam ($\lambda$=266 nm) on a classic Lloyd's mirror, a spin-coated photoresist (PR, Shipley 1813) layer is inscribed with 1D patterns. With proper exposure time, the PR is patterned with desired fill factor. Then, the c-Si layer is etched in a reactive-ion etcher with $CHF_3$ + $SF_6$ gas mixture. After residual PR removal, a Si 1D grating on quartz substrate is prepared as depicted in the second step of Fig. 6(a). To realize membrane-like structure, refractive index-matching oil (Cargille Lab. Series A) is placed on the Si grating. Thereafter, as seen in the fourth step, it is encapsulated with a bare quartz substrate. As seen in the photographic image, several 5 mm by 5 mm c-Si membrane cells are embedded in double quartz substrates. With the given large modulation strength ($\Delta\varepsilon = 3.48^2 - 1.45^2 = 10$), we designed these 1D membrane structures to exhibit new singular states. Figure 6(b) shows the $T_0$ spectra of a designed c-Si lattice for input angles $\theta$=0° and 2° where the grating parameter set ($\Lambda$=0.835 μm, $F$=0.522 and $H$=0.329 μm) is optimized by performing particle swarm optimization (PSO) algorithm with RCWA [46]. As illustrated in the inset, we consider the refraction angle of incident light between air ($n_a$=1) and quartz ($n_{Qz}$=1.45). At off-normal incidence $\theta$=2° under TE polarization, two a-GMRs appear as a singular state at $\lambda$=1.991 and EIT-like resonance at $\lambda$=1.373 μm. The resonant signatures of the singular state and EIT-like state are observed in the fabricated device as presented in Fig. 6(c). The grating parameter set of the fabricated c-Si membrane structure is ($\Lambda$=0.835 μm, $F$=0.396 and $H$=0.33 μm) with the greatest deviation from the design parameters being in the fill

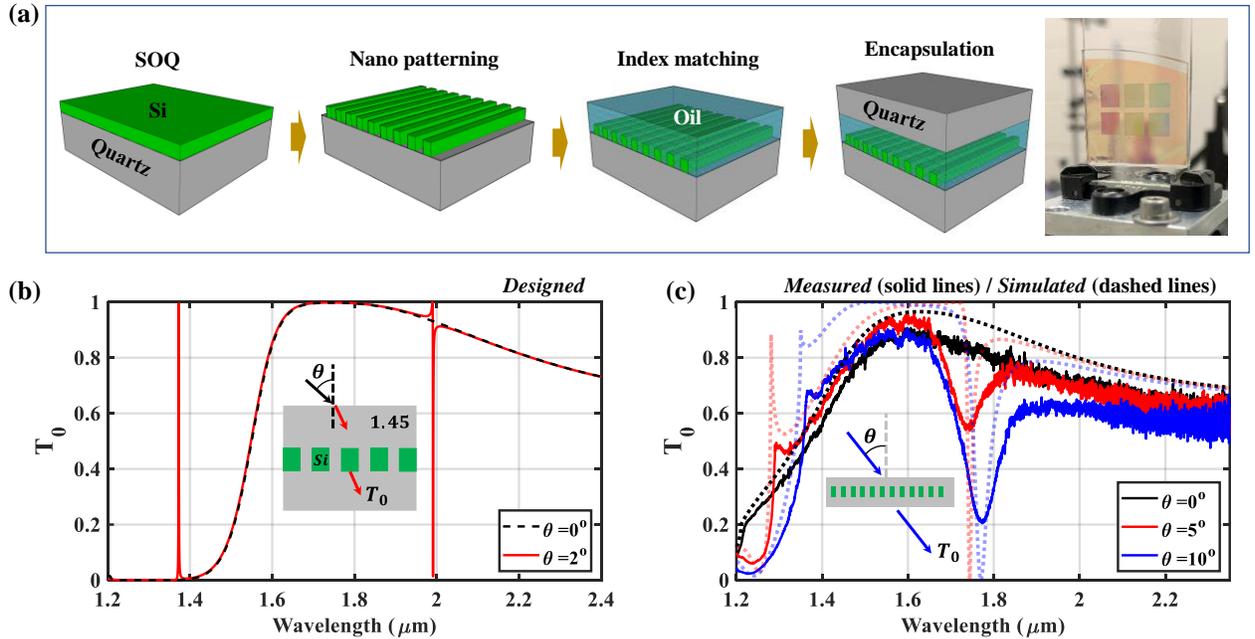

**Fig. 6.** Experimental verification of the existence of singular states in simple resonant lattices. (a) Fabrication procedure of a Si membrane that includes four steps: preparing an SOQ wafer (crystalline Si on quartz), nanopatterning, refractive-index matching, and encapsulating. A photographic image shows the membrane cells embedded in two quartz substrates. (b) Design of c-Si ($n_{Si}$=3.48) membrane in quartz ($n_{Qz}$=1.45) to expose a singular state where the grating parameter set ($\Lambda$=0.835 μm, $F$=0.522 and $H$=0.329 μm) is optimized by our PSO algorithm. In the $T_0$ spectra, the singular state and EIT-like resonance locate at $\lambda$=1.991 and 1.373 μm under oblique incidence ($\theta$=2°) in TE polarization. (c) Measured $T_0$ spectra (solid lines) for different incidence angles of $\theta$=0°, 5° and 10° where the parameter set of the fabricated device is ($\Lambda$=0.835 μm, $F$=0.396 and $H$=0.33 μm). For comparison, the simulated $T_0$ spectra (dashed lines) are depicted for the same grating parameters.

Factor F. To measure the $T_0$ spectrum, we use a near-IR spectrum analyzer (Yokogawa AQ6375) and supercontinuum light source (SuperK COMPACT, NKT Photonics). As represented in solid lines, the measured $T_0$ spectra show the singular state clearly albeit with large linewidths. The EIT-like a-GMR at the shorter wavelengths appears but somewhat weakly. Computing the spectra using the experimental parameters (dashed lines) shows a reasonably good match to the resonance positions of the measured $T_0$ spectra. Due to the divergence of the input Gaussian beam of the light source and imperfect fabrication, the bandwidth is broader and efficiency is lower relative to theory.

## 6. CONCLUSIONS

In summary, we have analytically and experimentally demonstrated singular states in simple optical lattices. These states appear as isolated, high-efficiency spectral lines arising out of a wideband low-reflectance floor and widely separated from other resonance features. In principle, these are perfectly reflecting BIC states under structural or angular asymmetric perturbation. Alternatively, they can be transmissive BICs within a high-reflectance wideband under analogy with "electromagnetically induced transparency." In past research, we established the applicability of Rytov's full effective-medium theory (EMT) to the physical description and design of resonant optical lattices [41]. There, we applied the symmetric formalism to reliably describe the behavior of various optical devices such as wideband reflectors, resonant bandpass filters, and guided-mode resonance polarizers. Here, we additionally show the utility of Rytov's asymmetric solution that has hitherto not been known to be useful. We find that this rejected Rytov analytical EMT solution predicts the dark BIC states essentially exactly for substantial lattice modulation levels. Thus, the established equivalent BIC homogeneous waveguide slabs possess propagation constants that represent the embedded eigenvalues foundational to the BIC state. The resonant lattices analyzed numerically represent membranes that approximate silicon nitride hosted in air. The attendant fabrication is challenging as large-area membranes in air with nanoscale thickness tend to buckle and break due to local stresses. Thus, we apply silicon-on-quartz wafers with refractive-index matching oil to retain an approximately uniform host medium. The fabricated devices demonstrate key effects predicted here with reasonable agreement between experimental data and numerical evaluation using the experimental final parameters. The work presented herein addresses unexplored concepts in resonant nanosystems thereby laying groundwork for new scientific discoveries and applications.

**Funding.** This research was supported, in part, by the UT System Texas Nanoelectronics Research Superiority Award funded by the State of Texas Emerging Technology Fund as well as by the Texas Instruments Distinguished University Chair in Nanoelectronics endowment.

**Acknowledgment**. Parts of this research were conducted in the UT Arlington Shimadzu Institute Nanotechnology Research Center. The authors thank Shin-Etsu Chemical Co, Ltd., Japan, for providing the SOQ wafers.

**Disclosures**. The authors declare no conflicts of interest.

**Data Availability**. Data underlying the results presented in this paper are not publicly available at this time but may be obtained from the authors upon reasonable request